# Manufacturing Process Optimization using Statistical Methodologies

Karthik Srinivasan[a]*; Amit Kumar[a]; Parameshwaran Iyer[a]; Abhinav Joshi[b]

[a]Robert Bosch Engineering and Business Solutions, India.
[b]Robert Bosch - Nashik Plant, India.

**Abstract:** Response Surface Methodology (RSM) introduced in the paper (Box & Wilson, 1951) explores the relationships between explanatory and response variables in complex settings and provides a framework to identify correct settings for the explanatory variables to yield the desired response. RSM involves setting up sequential experimental designs followed by application of elementary optimization methods to identify direction of improvement in response. In this paper, an application of RSM using a two-factor two-level Central Composite Design (CCD) is explained for a diesel engine nozzle manufacturing sub-process. The analysis shows that one of the factors has a significant influence in improving desired values of the response. The implementation of RSM is done using the DoE plug-in available in R software.

## 1. Background

Many a times, it is expensive to carry out random repeated experiments in a domain like manufacturing in order to carry out data collection and analysis. Design of Experiments (DoE) is a planned approach for determining cause and effect relationships applying statistical mathematical methods (Lazic, 2006). DoE is found to be highly applicable in almost all the scientific domains wherever controlled experiments can be done for analysis. RSM is defined in (Myers, Montgomery & Anderson-Cook, 2009) as a collection of statistical and mathematical techniques useful for developing, improving, and optimizing processes. Since RSM involves setting up an experimental design and using elementary optimization methods on the corresponding response surface plots, we can assume that RSM implicitly involves

---

* Corresponding Author. *Telephone*: +919448237388   |   *Email address:* karthik.srinivasan@live.com

DoE methodology and is useful to optimize the response with some given input factors. This paper explains formulation and implementation of RSM consisting of a simple 2-factor 2-level Central Composite Design using the open-source statistical software tool R. A manufacturing sub-process is considered as a use case for the methodology implementation.

The injector nozzle is a critical part in an engine sub assembly. It is manufactured in several phases and Robert Bosch is a leading player in nozzle manufacturing processes. The hydro-abrasive grinding process is a procedure in which a liquid with abrasive emulsion is passed through nozzle spray hole causing it to smoothen around the edges. The process quality is determined by amount of test oil flowing through it for a unit time. The objective of the study was to capture the influence of two input factors: abrasive liquid concentration and grinding pressure on output deviation from specified target.

## 2. Methodology and Formulation

RSM overall is a statistical technique for designing experiments, building models, evaluating the relative significance of several independent variables, and determining the optimum conditions for desirable responses (Demirel and Kayan, 2012).

A standard response surface plot of second order with two input factors and one response is expressed as in Equation 1.

$$Y = \beta_0 + \beta_1 X_1 + \beta_2 X_2 + \beta_3 X_1 X_2 + \beta_4 X_1^2 + \beta_5 X_2^2 + \varepsilon \tag{1}$$

Where, in Equation 1:

$\{\beta_1 X_1, \beta_2 X_2\}$ are the first order effects;

$\{\beta_3 X_1 X_2\}$ is the interaction effect and;

$\{\beta_4 X_1^2, \beta_5 X_2^2\}$ are the quadratic effects.

The response surface can be considered as a first order relation between the outcome and factor inputs if the second order effects are not significant. In the second order effects, either the interaction effect or the quadratic effect or both can be included in the response surface function. This simple representation in steps of order of the function makes it easy to interpret in terms of different kinds of plots and capture non-linear quadratic effects if any.

In the example considered, the two input factors as well as the outputs are quantitative and the initial starting points for analysis were known based on the experience of domain experts. It is required to derive causal inferences of the two factors on the deviation from target value of outcome for process optimization in terms of minimum absolute deviation/error. The sub-process involves a lot of other process parameters. From a domain perspective two important factors affecting the response were identified. Hence it is required to evaluate the change in output in response to increasing or decreasing the two inputs without varying other parameters.

The two most common designs extensively used in RSM are CCD and the Box-Behnken design (BBD). The CCD is ideal for sequential experimentation and allows a reasonable amount of information for testing lack of fit while not involving an unusually large number of design points (Demirel and Kayan, 2012). Since only two factors were considered and both of them were quantitative, a full-factorial experiment with CCD was selected for constructing the response surface.

In the current study, it was required to study the effect of nozzles coming from different streams of the previous assembly sub-process. An additional indicator term was included for treatment of these two different blocks of data. In this way, the inference using sum of squares between blocks similar to analysis of variance could be incorporated in the response surface function.

The components of a full CCD suggested in the paper (Lundstedt et. al, 1998) are:

1. A full factorial or fractional factorial design.
2. Experiments at the centre
3. Experiments where input levels = +/- α. These points are situated on the axis in a co-ordinate system with distance +/- α from the origin; known as *axial* points.

A design is said to be rotatable if the variance is a function of only distance of input vector from the origin, not its direction or the type of input (Park, Lim & Baba, 1993). Thus, when a design is rotatable, the variance of predicted values is the same at all possible input vectors that are equidistant from the design centre. The axial points were introduced in CCD to include the property of rotatability in the response surface. Considering axial points increases number of experiment points and increases cost if rotatability is not highly desired. Moreover, the input factors could not be varied more than due to limited resolution and process sensitivity to change in inputs. Hence in this paper, only the first two components of CCD are considered as the optional third component was not suitable for the analysis.

In the final experimental design, two blocks were considered with two levels and a center value and eight repetitions were done and outcomes were recorded. The schema for the design is given in Table 1.

Table 1: Experimental schema for use-case

| Run order | Factor Input 1 | Factor Input 2 | Outcome (to be measured) |
|---|---|---|---|
| 1 | Low(-) | Low(-) | |
| 2 | Low(-) | High(+) | |
| 3 | High(+) | Low(-) | |
| 4 | High(+) | High(+) | |
| 5 | Center value | Center value | |

The final expression for the response surface in this study is given in Equation 2.

$$Y_{Error} = \beta_0 + \beta_1 X_{liquid} + \beta_2 X_{pressure} + \beta_3 X_{liquid} X_{pressure} \\ + \beta_4 X_{liquid}^2 + \beta_5 X_{pressure}^2 + \beta_6 I_{Block} + \varepsilon \quad (2)$$

Where, in Equation 2:

$Y_{Absolute\_error}$ is the absolute error or absolute deviation from target value

$X_{liquid}$ is the input factor of abrasive liquid concentration

$X_{pressure}$ is the grinding pressure

$I_{Block}$ is the qualitative indicator specifying block {1,2}

{$\beta_1 X_{liquid}$, $\beta_2 X_{pressure}$} are the first order effects

{$\beta_3 X_{liquid} X_{pressure}$} is the interaction effect and

{$\beta_4 X_{liquid}^2$, $\beta_5 X_{pressure}^2$} are the quadratic effects

## 3. Analysis using R software

R is a language and environment for statistical computing and graphics and is available as free software under the terms of General Public License (GNU). It has various add-on packages of code and additional plug-ins specific for different applications. Apart from the standard packages required for standard computation and graphics, package *rsm* with its dependencies written by (Lenth, 2010) is useful for the formulation of RSM. Additionally, (Grömping, 2011) has developed a flexible and easy to use DoE plug-in called *RcmdrPlugin* which is installed as an add-on over the basic-statistics graphical user interface R commander by (Fox, 2005). The packages with the dependencies can be downloaded using the *install.packages* command in R software.

## 4. Results

The response plot can be analyzed to (a) Find the effects of significant input factors on response as well as (b) Determine the best direction for process improvement or arrive at an optimal operating condition. Optimization using steepest descent method produces a path that results in a maximum decrease in response (Fletcher, Powell and MJ, 1963). The steepest descent can be easily visualized in a two factor response plot as the perpendicular direction to the contour lines in the contour plot shown in Figure 1. The slope of the perspective plot as given in Figure 2 also gives an understanding of the overall response with respect to the significant effects of input factors.

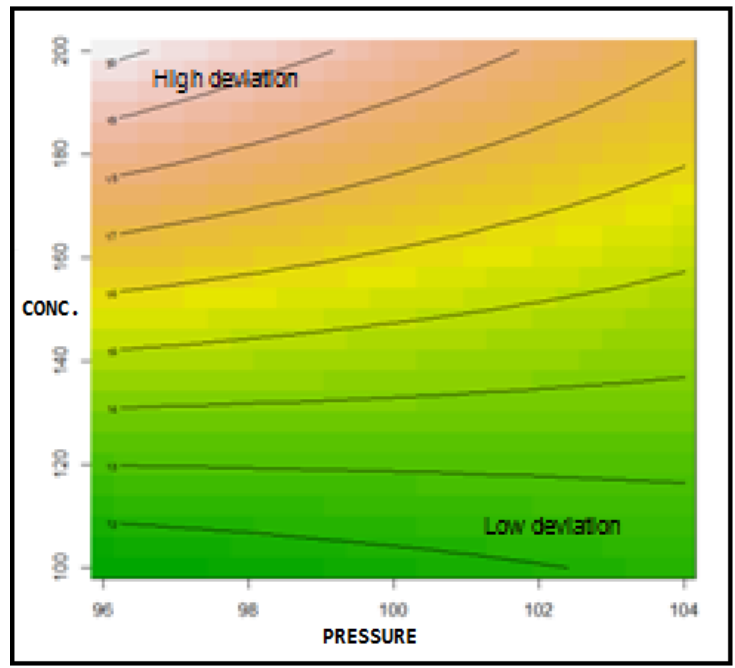

Figure 1: Contour Plot showing absolute error levels across input space

The statistical analysis of response plot is not presented in this paper but confirmed (a) significant difference between the two experimental blocks initially considered and (b) Level of Abrasive liquid concentration had a positive effect on absolute error. The local minima was reached at $(X_{liquid}, X_{pressure})$ = (Low_value, Low_value) which can be verified by observing the bottom left point in the perspective plot having minimum absolute error. It was found that increase in abrasive liquid concentration increased the absolute error in the process.

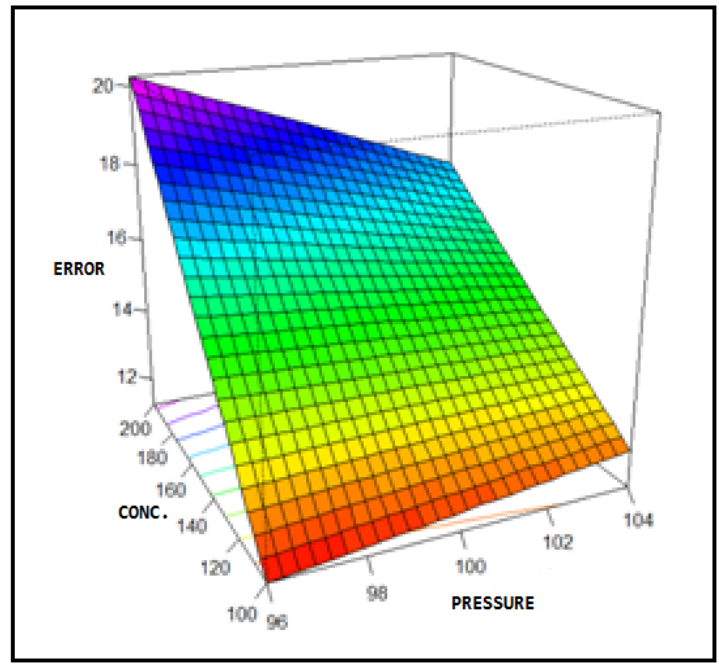

Figure 2: Perspective Plot of Response surface

## 5. Discussions

The data analytical problems in the manufacturing sector are often characterized by small to mid-size datasets with lot of captured variables and inherent un-captured parameters. Isolating effects of parameters on process outcome is not trivial but a certain level of controlled environment for selective empirical studies can be achieved. It is often observed that, trial and error methods are adopted to discover any useful inference about parameter-outcome relationships. These methods are often successful based on user intuition and experience in monitoring the process. Various statistical methods and descriptive analysis are used and aid in decision making. Even though, DoE and RSM is quite prevalent in the research fraternity, it has a lot of scope of application to aid in process improvement in an easy, effective and systematic manner. A simple analysis such as the one considered in the use case, not only gives important insights but also gives a direction for identifying scope for further such analysis. Open source tools, packages and plug-ins such as the ones discussed in the paper are free, easily accessible and have online technical support assistance through web discussion forum. The availability of plug-ins such as *RcmdrPlugin* has made the DoE application to problems all the more easier and user-friendly which should encourage more extensive usage of RSM in the future.

## Disclaimer

The objective of this paper is towards explaining the application of RSM and consequent statistical analysis to optimize manufacturing processes. The use case is intentionally devoid of any kind of specific knowledge about the Hydro-Abrasive process.